\documentclass[a4paper]{mem}
\usepackage{natbib}
\usepackage{graphicx}
\usepackage[a4paper]{hyperref}
\idline{3}{1}
\begin{document}

\title{Blazars: the next gamma-ray view of GLAST}

\author{Stefano Ciprini\inst{1,2}}

\institute{Physics Department and Astronomical Observatory,
University of Perugia, via Pascoli, 06123 Perugia, Italy\\ \and
INFN, Perugia Section, via Pascoli, 06123 Perugia, Italy\\
~~\email{stefano.ciprini@pg.infn.it}}

\authorrunning{Ciprini}

\titlerunning{Blazars: the next gamma-ray view of GLAST}

\abstract{Blazars, the extreme family of AGN, can be strong
gamma-ray emitters and constitute the largest fraction of
identified point sources of EGRET. The next Gamma-ray Large Area
Space Telescope (GLAST) is a high energy (30MeV-300GeV) gamma-ray
astronomy mission, planned for launch at the end of 2006. GLAST
performances will allow to detect few thousands of gamma-ray
blazars, with a broad band coverage and temporal resolution, also
in quiescent emission phases, providing probably many answers
about these sources.
   }

   \maketitle

%
\section{The Italian contribution to the Large Area Telescope of GLAST}
\begin{figure*}[t!]
   \centering
   \resizebox{12cm}{!}{\rotatebox[]{-90}{\includegraphics{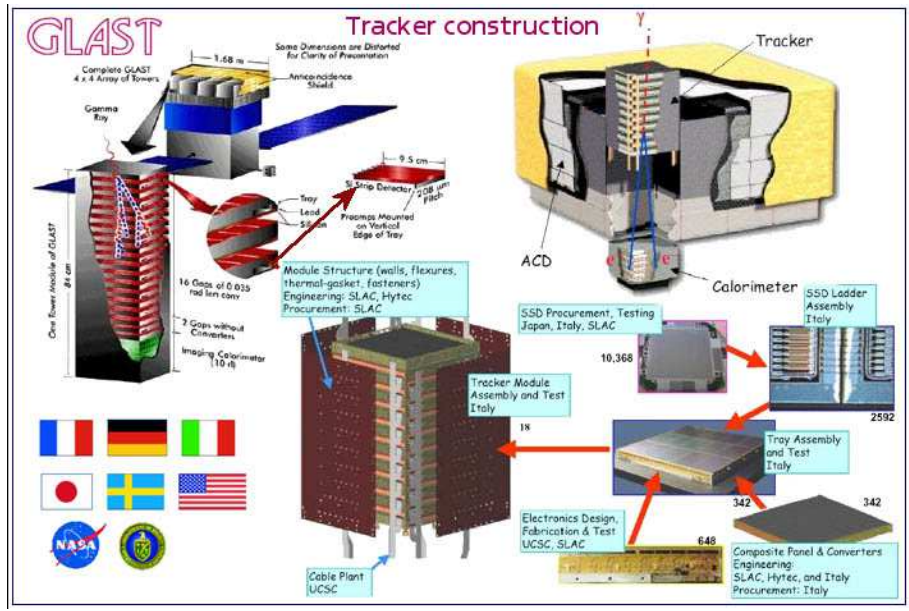}}}
   \vskip -2cm
   \caption{Scheme of the Large Area Telescope (LAT) of GLAST, formed by an array of 16
   identical ``tower'' modules (+2 for test beam), each with a tracker (silicon micro-strips vertex detectors)
   a calorimeter (CsI with PIN diode readout), and DAQ module. The array is surrounded by
   finely segmented ACD (plastic scintillator with PMT readout). A very big amount of silicon
   microstrips will be used for the tracker, equivalent to a total surface of 83m$^2$
   of silicon detector (11500 SSDs).
   }
              \label{fig:tracker}\vskip -0.3cm
    \end{figure*}

The Gamma-ray Large Area Space
Telescope\footnote{\texttt{http://glast.gsfc.nasa.gov/}\\and~\
\texttt{http://www-glast.stanford.edu/}} (GLAST) project is part
of NASA's SEU Program, within the Office of Space and Science,
funded and realized with the collaboration of NASA, U.S.
Department of Energy, institutions and government agencies in
France, Germany, Japan, Italy and Sweden. GLAST is a next
generation high-energy gamma-ray observatory, designed for making
observations of astronomical gamma-ray sources in the energy band
extending from 30 MeV to 300 GeV. It follows in the footsteps of
the CGRO-EGRET
mission\footnote{\texttt{http://cossc.gsfc.nasa.gov/egret/}}
(operational between 1991 and 1999), and its launch is scheduled
for the end of 2006. GLAST will have two scientific instruments:
(1) the Large Area Telescope (LAT), an imaging, wide field-of-view
telescope (composed of a tracker based on silicon micro-strip
vertex detectors and a calorimeter), sensitive to gamma-rays over
the energy range from $\sim$20-30 MeV to more than 300 GeV, and
(2) the Burst Monitor (GBM), sensitive to transient bursts from 10
keV to 25 MeV. The LAT (see Fig. \ref{fig:tracker}) is a
pair-conversion telescope, formed by 16 ``tower'' modules, each
with a tracker based on silicon microstrips (EGRET was based on
gas spark chambers), a calorimeter (CsI with PIN diode readout)
and DAQ module. The array is surrounded by finely segmented Anti
Coincidence Detectors (ACD, plastic scintillator with PMT
readout). The amount of silicon strip detector used for the
tracker is impressive for a space-borne project: it is equivalent
to a surface of 83m$^2$ (about 11500 Single Strips Detectors
SSDs), approximatively the same used in the next ATLAS experiment
at CERN \citep[for a description of the LAT see for
example][]{michelson03,bellazzini02}. Recent descriptions of some
scientific topics and software for GLAST can be found in
\citep{ciprini03gammaproc}.
%
%
%
\par Funding and participating Italian institution to GLAST, is
mainly \textit{Istituto Nazionale di Fisica Nucleare} (INFN, Bari,
Padova, Perugia, Pisa, Rome2 and Trieste-Udine sections) for the
LAT project, joined to the Italian Space Agency (ASI) and IASF-CNR
of Milan. ASI is preparing a compact $\gamma$-ray mission at this
energy bands
AGILE\footnote{\texttt{http://agile.mi.iasf.cnr.it/}}, which will
anticipate GLAST and then cooperate overlapping with it. Moreover
Italy is a traditional partner for important NASA astrophysics
exploration missions (Swift, Constellation X, JWST etc.). In the
case of GLAST, the participation also of INFN to construction and
testing, will permit an astrophysics and particle physics
partnership. In this view the project is open to the Italian
astronomical community.
%
%
   \begin{figure}[t!]
   \centering
   \resizebox{4.5cm}{!}{\rotatebox[]{0}{\includegraphics{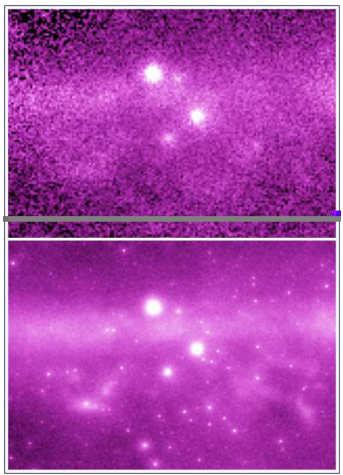}}}
   \vskip -0.2cm
   \caption{EGRET observed (upper panel, CGRO Phases 1-5) and GLAST-LAT simulated (lower panel, one-year
   all sky survey) views of the Galactic Anticenter in gamma-rays above 100
   MeV. Field is about 80 degree x 54 degree. (Reproduced with the permission of S. Digel, SLAC).}
              \label{fig:anticentercomp} \vskip -0.4cm
    \end{figure}
%
%
\section{GLAST and gamma-ray blazars}
%
The GLAST LAT is a considerable improvement over its successful
predecessor EGRET, with its broad energy range (0.03-300GeV), the
10000cm$^2$ of effective area (at 10 GeV), the 9\% of energy
resolution (at 0.1-100 GeV), with 2.4sterad of FOV, and an angular
resolution of 3.4$^{\circ}$ at 100 MeV, 0.086$^{\circ}$ at 10 GeV.
This gives a point source sensitivity (at 5$\sigma$) above 100
MeV, better than $3\times 10^{-9}$ photons cm$^{-2}$ s$^{-1}$, for
the first year of all-sky observing mode, at high Galactic
latitude \textit{b}, and for sources with $E^{-2}$ photon spectra
\citep{digel03}, 20-30 times better than EGRET. Source location
determination is 0.4' at 1$\sigma$ radius, (source flux 10$^{-7}$
photons cm$^{-2}$ s$^{-1}$, $E>100$MeV.
%
%
   \begin{figure}[t!]
   \centering
   \resizebox{\hsize}{!}{\rotatebox[]{0}{\includegraphics{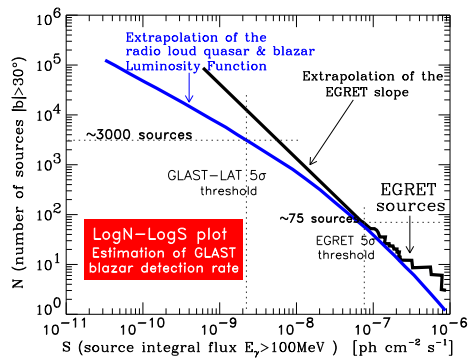}}}
   \vskip -0.2truecm
   \caption{LogN-LogS plot with the estimated GLAST detection
   rate of blazars. In black the curve of the EGRET observations and the extrapolated slope.
   The blue(gray) curved line is an extrapolation
   based on the radio-loud quasars-blazars Luminosity Function
   \citep[after ][]{stecker96,gehrels99}.}
              \label{fig:logNlogS}%
    \end{figure}
%
In Fig. \ref{fig:anticentercomp} is showed a comparison of the
Anticenter region at $E > 100$MeV, observed by EGRET, and the
expected GLAST-LAT view. The PSF and effective area are still
under calculation, but Fig. \ref{fig:anticentercomp} provide an
adequate and correct visual impression of the quality of LAT data,
which will permit a strong improvement in our knowledge on the
$\gamma$-ray sky, \citep[for a topics review see, e.g. the
\textit{GLAST science document}, ][]{michelson01}.
\par One of the main research fields for GLAST are the AGN,
especially blazars (flat spectrum radio quasars, FSRQs, and BL Lac
objects), for number of sources, and data quality. Gamma-rays
provide a precious probe to investigate the extreme physical
conditions in blazars jets, at sub-parsec scales and relativistic
regimes. Over 3000 $\gamma$-ray blazars, are expected to be
detected \citep{stecker96,gehrels99} (prudential number based on
the LogN-LogS relation assuming, a luminosity function
proportional to that of radio-loud quasars--blazars, Fig.
\ref{fig:logNlogS}). The GLAST--LAT will measure the continuum
spectral energy distribution (SED) of blazars, in uncovered
$\gamma$-ray energy bands (overlapping in the higher energy tail
with ground--based Cherenkov telescopes). GLAST with its broand
band coverage and temporal resolution will alow to identify and to
constrain leptonic (SSC, EC, pairs) and hadronic ($\pi^{0}$ decay,
proton) emission processes. It will be able to track $\gamma$-ray
flares and variability, to correlate $\gamma$-ray emission with
simultaneous multiwavelength observations, and will investigate
the relations between $\gamma$-ray flares and the VLBI
superluminal radio components and plasma blobs, shedding light on
the disk-jet connection and on the nuclear activity. Moreover it
will probe the extragalactic background light (EBL) through the
absorption of $\gamma$-rays from blazars at higher $z$. The rate
of the blazar flare emission detectable with GLAST, should be
favoured at energies around 100 MeV \citep{dermer02}.
%
%
%
   \begin{figure*}[t!]
   \centering
   \resizebox{13cm}{!}{\rotatebox[]{0}{\includegraphics{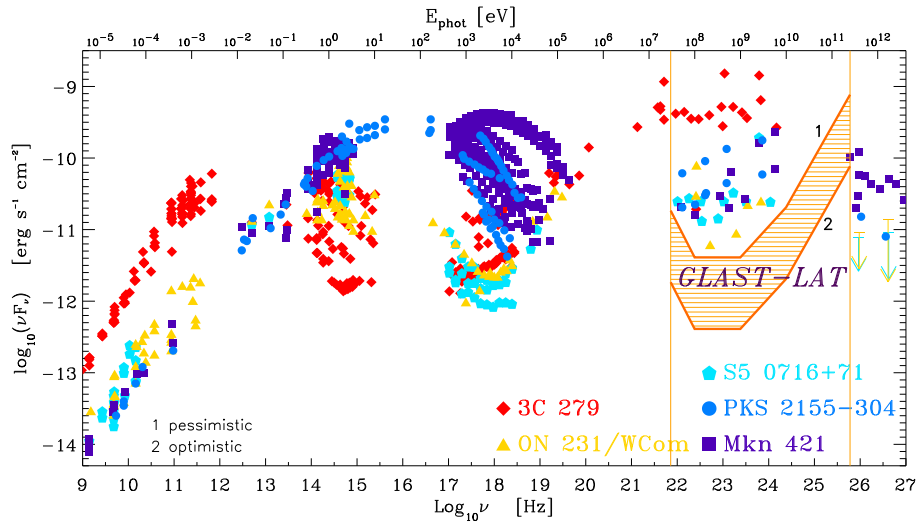}}}
   \vskip -0.3cm
   \caption{Overall SEDs of some blazars,
   and a qualitative representation of the expected GLAST Large Area Telescope
   sensitivity (prudential estimations based on 80 hours at 5$\sigma$,
   and mean $dN(E)/dE \propto E^{-2}$). The limits are only qualitative because
   response functions (i.e. effective area, PSF, energy resolution etc.) are
   under calculation. SED data refers to different epochs (error
   bars not reported for clarity).}
              \label{fig:sedandglast} \vskip -0.2cm
    \end{figure*}
%
%
\par The SED of blazars show a two-bump
structure, produced by synchrotron emission and inverse Compton
(IC) scattering of soft photons in leptonic descriptions (see Fig.
\ref{fig:sedandglast}). In the synchrotron self-Compton scenario
(SSC), the diffusive shock acceleration of electrons within a
relativistic jet pointing toward the observer, produces
synchrotron radiation, which is up-scattered by IC by the same
relativistic particles population. This description seems to
account well for the emission of the HBL group (e.g. PKS 2155-304
and Mkn 421 in Fig. \ref{fig:sedandglast}), that are usually also
TeV emitters. On the other hand, in the leptonic scenario, FSRQs
needs of thermal components and external--jet seed photons
(external IC descriptions), to explain the relevant gamma--ray
dominance (e.g. 3C 279 in Fig. \ref{fig:sedandglast}). In the
picture is sketched the qualitative sensitivity of the LAT. During
flaring GLAST will be able to track the spectral evolution of the
IC bump (MeV-GeV peaked), as for the synchrotron one (peaked in
the IR-soft-X range), detecting also the quiescent emission of
blazars. This will remove the current degeneracy in theoretical
models, providing strong constraints.
%
%


\end{document}